\documentclass[twocolumn,preprintnumbers,a4paper,10pt,unsortedaddress,groupedaddress,superscriptaddress]{revtex4-1}
\usepackage{graphicx}
\usepackage{amssymb}
\usepackage{amsmath}
\usepackage{dcolumn}
\usepackage{bm}
\usepackage{color}
\usepackage[english]{babel}
\usepackage[novolumeabbr]{unitsdef}
\newunit{\dBm}{dBm}
\newunit{\dB}{dB}

\begin{document}

\title{Nonlinear Induction Detection of Electron Spin Resonance}
\author{Gil Bachar}
\email{gil@tx.technion.ac.il}
\affiliation{Department of Electrical Engineering, Technion, Haifa 32000 Israel}
\author{Oren Suchoi}
\affiliation{Department of Electrical Engineering, Technion, Haifa 32000 Israel}
\author{Oleg Shtempluck}
\affiliation{Department of Electrical Engineering, Technion, Haifa 32000 Israel}
\author{Aharon Blank}
\affiliation{Schulich Faculty of Chemistry, Technion, Haifa 32000 Israel}
\author{Eyal Buks}
\affiliation{Department of Electrical Engineering, Technion, Haifa 32000 Israel}

\begin{abstract}
We present a new approach to the induction detection of electron spin resonance (ESR) signals exploiting the nonlinear properties of a superconducting resonator. Our experiments employ a yttrium barium copper oxide (YBCO) superconducting stripline microwave (MW) resonator integrated with a microbridge. A strong nonlinear response of the resonator is thermally activated in the microbridge when exceeding a threshold in the injected MW power. The responsivity factor characterizing the ESR-induced change in the system's output signal is about 100 times larger when operating the resonator near the instability threshold, compared to the value obtained in the linear regime of operation. Preliminary experimental results, together with a theoretical model of this phenomenon are presented. Under appropriate conditions nonlinear induction detection of ESR can potentially improve upon the current capabilities of conventional linear induction detection ESR.
\end{abstract} 

\maketitle

Electron spin resonance (ESR) is a well-known method enabling direct measurements of the electron spin Hamiltonian, with applications ranging from biology to materials science and physics \cite{Poole_ESR, Misra_EPR, Abragam_NMR}. However, a significant drawback of conventional ESR is its relatively low sensitivity compared with other spectroscopic and analytic techniques, such as fluorescence and mass spectrometry. For example, the world record in electron spin sensitivity stands today at $\sim 10^6$ spins per 1 sec of acquisition (often denoted as spins/$\sqrt{\hertz}$) or slightly more than $10^4$ spins in a reasonable $\sim 1$ h of acquisition \cite{Twig_076105}, which is still far from single electron spin sensitivity. This current sensitivity limitation also restricts the available imaging resolution of heterogeneous samples. Thus, while the laws of physics do not set a limit to the spatial resolution of ESR (at least up to the atomic-length scale), in practice, as the image's voxel (volumetric pixel) size decreases, it contains less and less spins and thus quickly runs into the sensitivity limitation wall. For example, the systems in our laboratory, achieved recently a $440\nm$ resolution, limited mostly by spin sensitivity \cite{Shtirberg_043708}.

The above-mentioned numbers for sensitivity and resolution refer to ESR systems that employ "induction detection", namely, they make use of Faraday's law for the detection of ESR signals by means of a pick-up coil or a microwave (MW) resonator. Induction detection is the basic principle behind all commercial state-of-the-art ESR systems; it enables the acquisition of high resolution spectroscopic data with complex pulse sequences; facilitates the use of efficient imagining methodologies (meaning that signals are acquired and averaged in a parallel fashion from the entire sample); and features convenient sample handling. While our work is focused on induction detection ESR, other groups have looked into alternative detection methods in an attempt to increase sensitivity and resolution. These include, for example, magnetic resonance force microscopy \cite{Rugar_329}, scanning tunneling microscopy ESR \cite{Durkan_458}, spin-polarized STM \cite{Meier_82}, electrically detected magnetic resonance \cite{Harneit_216601}, and indirect spin detection via diamond nitrogen-vacancy (NV) centers \cite{Grotz_055004}. While these new techniques may improve even more in the future, they have some inherent limitations, resulting in limited applicability.

It is therefore evident that there is still a strong need to greatly improve the sensitivity of induction detection ESR up to the ultimate single spin sensitivity (in a reasonable acquisition time of $\sim 1$ h or less), making it a generally applicable method for noninvasive detection and imaging of small numbers of electron spins.
Here we take a step in that direction and show that the sensitivity of induction detection may be enhanced if it is employed in conjunction with a unique new class of non linear superconducting resonators.  In our new scheme, the sample's resonance properties affect the non linear properties of the resonator, thereby resulting in a complete new approach to the detection of ESR signals.
Our initial experimental results demonstrate this new approach in practice and are accompanied by a theoretical analysis explaining our observations. The measured responsivity of our resonator is enhanced by a factor of up to 100 when operating the system in the nonlinear regime.

\begin{figure} [b]
\includegraphics[width=\columnwidth]{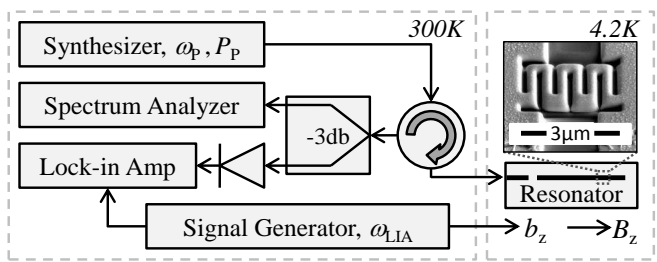}
\caption{Experimental setup for ESR signal measurements. The device is installed in a cryostat where a strong static magnetic field $B_z$ is applied. A single coherent tone, with angular frequency $\omega_\textrm{P}$ and power $P_\textrm{P}$, is injected into the feedline. The reflected signal is split and measured by a spectrum analyzer in the frequency domain, and by a lock-in amplifier (LIA). The LIA is tuned to the frequency of the slow magnetic field modulation $b_z \cos(\omega_\mathrm{LIA} t)$ that is applied through a wire. The inset shows an electron micrograph of the microbridge.}
\label{Fig:ExperimentalSetup}
\end{figure}

The experimental setup is schematically presented in Fig. \ref{Fig:ExperimentalSetup}. It is composed of a stripline waveguide MW resonator \cite{Pozar_MWE}, with a characteristic impedance of $50 \ohm$. Here we employed its second mode at $6.1\GHz$. The resonator is weakly capacitively coupled to a feedline. In the resonator, a narrow section with a length of $11.5 \micm$ and width of $0.3 \micm$ is defined. This section is referred hereafter as the "microbridge".

We began the fabrication process with a 0.5-mm-thick sapphire wafer coated with $150\nm$ of yttrium barium copper oxide ($\textrm{YBCO}$), and a gold cup layer of $200 \nm$. 
Gold contacts and alignment marks were patterned using standard photolithography and wet etching. The waveguide was patterned with electron beam lithography and wet etching. In the final step, the microbridge was patterned using a focused ion beam (FIB) system \cite{Blank_2786}.

The resonator was covered with another $0.5 \mm$ thick sapphire wafer. It was then packed in a copper box whose inner top and bottom surfaces are covered with niobium to provide superconducting ground planes for the stripline structure. Before installation, a 1-mm-diameter hole was drilled on the top sapphire wafer, just above the microbridge. The hole was then filled with a common stable free radical powder (DPPH from Sigma-Aldrich). In addition a superconducting wire was placed between the two sapphire wafers, below the hole, and perpendicularly to the stripline. The wire allows applying low-frequency magnetic field modulation to the paramagnetic sample.

The device was fully immersed in liquid helium inside the core of a superconducting coil. The ESR signal was measured in a continuous wave (CW) mode, similar to any conventional ESR system, with the exception that automatic frequency control was not used to track the resonator's resonance frequency. A monochromatic MW signal at a frequency close to the resonance frequency was injected. The reflected signal was measured while the external magnetic field was slowly scanned. In another setup, the resonator $S_{11}$ reflection coefficient was measured using a network analyzer. The resonator's loaded Q factor was found to be on the order of 5000.

\begin{figure} [b]
\includegraphics[width=\columnwidth]{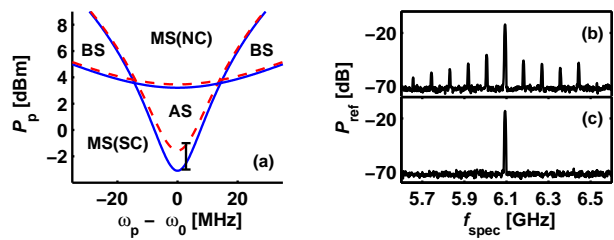}
\caption{(color online) Panel A: Self modulation stability diagram. The stability map, which is obtained by Eq. (\ref{Eq:Borderlines}) contains 4 stability zones: the superconducting mono stable zone, MS(SC), the normal conducting mono stable zone, MS(NC), the bi stable zone, BS, and the astable zone, AS.
We compare the case were $B_z$ is far from $B_z^\mathrm{ESR}$ (blue-solid lines) and the case of $B_z = B_z^\mathrm{ESR}$ (red-dashed lines). The experiment in Fig. \ref{Fig:Lockin} is done along the black line, at input power range $-3\dBm < P_\mathrm{p} < -1\dBm$, and slightly above the resonance frequency. Panels b-c: Data from spectrum analyzer at the monostable (C, $P_\mathrm{p} = -3\dBm$) and  astable (b, $P_\mathrm{p} = -1\dBm$) zones, when $B_z$ is far from $B_z^\mathrm{ESR}$.}
\label{Fig:StabilityDiagram}
\end{figure}

In previous work we have shown that the microbridge-integrated resonator system has a non linear response to a coherent tone injected close to the resonator's resonance frequency \cite{Segev_096206}. When input power exceeds a certain threshold, the power reflected off the resonator shows a self-excited amplitude modulation. We will refer to this threshold hereafter as the "modulation threshold". In Fig. \ref{Fig:StabilityDiagram} the spectral density of the reflected signal is demonstrated, above (panel b) and below (panel c) the modulation threshold. As will be shown below, the ESR responsivity can be enhanced when operating the resonator close to the modulation threshold.

\begin{figure} [b]
\includegraphics[width=\columnwidth]{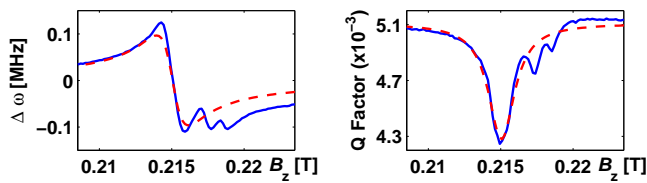}
\caption{(color online) The resonator's resonance frequency and Q factor extracted from $S_{11}$ measurements under $B_z$ scan. Experimental results (solid blue line) are compared with theoretical prediction based on Eq. (\ref{Eq:DeltaOmegaAndQ}) (dashed red line). The parameters used in Eq. (\ref{Eq:ChiPrime}) are: $t_1=t_2=1/61\MHz/\pi$, $N = 2\cdot 10^{27} \mathrm{spins}/\meter^{3}$ \cite{Yalcin_094105}, $T_\mathrm{ESR}=4.2\kelvin$, and $\eta = 8.6\cdot 10^{-5}$. The simulation takes into consideration a single homogenous broadened ESR line, while in experimental data g-factor anisotropy and hyperfine effects are observed \cite{Lundi_505} .}
\label{Fig:S11}
\end{figure}

We began by examining the linear response of the resonator to the ESR signal. The resonator was excited with power well below the modulation threshold, while the static magnetic field was slowly scanned (0.01\tesla/\minute). We measured the reflection coefficient of the resonator as a function of the input frequency. Near a magnetic field of $B_z^\mathrm{ESR}=0.215 \tesla$ the resonance frequency of the resonator was shifted and the Q-factor decreased (see Fig. \ref{Fig:S11}).
The results obtained by us are similar to those presented by others dealing in the coupling between a paramagnetic sample and a linear superconducting resonator \cite{Bushev_060501, Schuster_140501, Staudt_arxiv}.

\begin{figure} [b]
\includegraphics[width=\columnwidth]{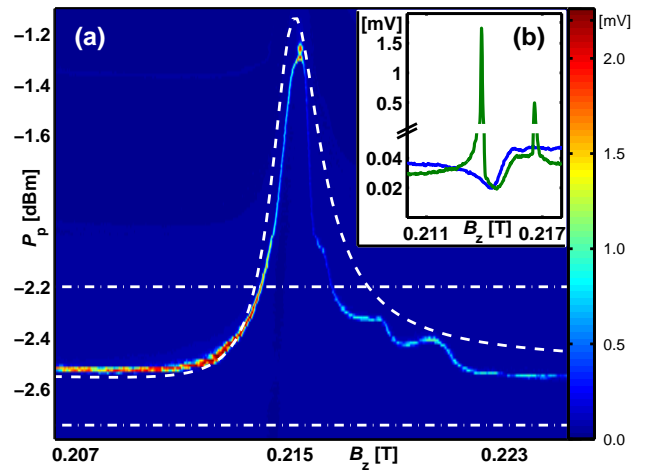}
\caption{(color online) ESR spectrum obtained in the linear and non-linear regimes. (A) The signal reflected off the resonator, as a function of the static magnetic field and the input power, measured using a LIA at $1.23 \kilohertz$. The ESR spectrum can be obtained from the non-linear response readings (local maxima of the graph), and is compared to analytical results obtained from Eqs. (\ref{Eq:SuperBorderline}) and (\ref{Eq:DeltaOmegaAndQ}) (dashed white line). The fitting parameters are $H = 7 \mu \watt/\kelvin$ and $T_c-T_0 = 70\kelvin$. (B) Cuts from the two dotted lines which are marked in the colormap: $-2.8 \dBm$ (solid-blue) and $-2.2 \dBm$ (solid-green) input power, showing linear and non linear response respectively. Note the discontinuity in the vertical axis, which was created in order to make visible the relatively small change in the linear regime.}
\label{Fig:Lockin}
\end{figure}

We then examined the response of the resonator at the non linear regime. We excited the resonator by injecting a monochromatic tone at frequency $\omega_\mathrm{p}$ which is close to the resonance frequency, and power in the vicinity of the modulation threshold. The static magnetic field was slowly ramped around $B_z^\mathrm{ESR}$, while
a low frequency ($1.23 \kHz$) low amplitude ($20 \micro \ampere$) ac-current was injected into the wire (as in conventional CW induction detection schemes) providing the magnetic field modulation. A lock-in amplifier (LIA) was tuned to the modulation frequency and measured the envelope of the signal reflected off the resonator. Slightly below the modulation threshold, the response was linear and power reflected off the resonator was proportional to the $S_{11}$ parameter (the reflection coefficient); therefore the LIA measured the derivative of the $S_{11}$ parameter with respect to the magnetic field. The ESR-induced change in the LIA signal was seen at the same value of $B_z^\mathrm{ESR}$ as in direct $S_{11}$ measurement (Fig. \ref{Fig:Lockin}-b, blue line). When the input power was set exactly at the modulation threshold, the LIA signal increased significantly by a factor of up to 100 (Fig \ref{Fig:Lockin}-b, green line). As we scanned the static magnetic field we found that the non linearity power threshold shifted; at $B_z^\mathrm{ESR}$ the power threshold increased by $\sim 1.2 \dB$ relative to the case where the static field is far from the resonance value. Thus, by scanning over the $P_\mathrm{p}$ - $B_z$ plane, and measuring the LIA signal we can obtain the ESR spectrum. (see Fig. \ref{Fig:Lockin}-a).

To analyze the results, we consider an ESR sample placed near a current anti node in a stripline MW resonator embedded with a microbridge. An external static magnetic field tunes the Larmor frequency of the ESR sample close to the resonator's  resonance frequency. The dynamics of the resonator in such a case can be captured by two coupled equations of motion. The full derivation is provided in \cite{Segev_096206}. Below we describe them in brief.

The resonator is driven by a coherent tone $a^{in}e^{-i\omega_\mathrm{p}t}$ injected into the test port, which is weakly coupled to the resonator, where $a^{in}$ is a constant amplitude and $\omega_\mathrm{p}$ is the drive angular frequency. The mode amplitude $A$ can be written as $A=\hat A e^{-i\omega_\mathrm{p}t}$, where $\hat A \left( t \right)$ is a complex amplitude, which is assumed to vary slowly on the time scale of $1/\omega_\mathrm{p}$. In this approximation, the equation of motion of $\hat A$ reads
\begin{equation}
\frac{\mathrm{d} \hat A}{\mathrm{d}t}=\left[  i\left(  \omega_\mathrm{p}-\omega_{0}\right)  -\gamma\right] \hat A-i\sqrt{2\gamma_{1}}a^{in}+c^{in},
\label{Eq:dA/dt}
\end{equation}
where $\omega_{0}$ is the resonance frequency, $\gamma=\gamma_{1}+\gamma_{2}+\gamma_{3}$, $\gamma_{1}$ is the coupling constant between the resonator and the feedline, and $\gamma_{2}+\gamma_{3}$ is the damping rate of the mode, where $\gamma_{2}$ denotes the dumping rate of the microbridge, and $\gamma_{3}$ is the dumping rate of all other loss mechanisms. The resonator's $Q$-factor is defined as $Q=\omega_\mathrm{0}/\gamma$.
The term $c^{in}$ represents a random-phase input noise with correlation function: $\langle c^{in}\rangle=0$ and $\langle c^{in}(t)c^{in\ast}(t^{\prime})\rangle = 2\gamma \frac{k_\mathrm{B}T}{\hbar\omega_{0}} \delta\left(t-t^{\prime}\right)$,
for the case of thermal equilibrium at high temperature ($k_\mathrm{B}T \gg \hbar\omega_{0}$), where $k_\mathrm{B}$ is Boltzmann's constant.

We consider the case where non linearity is generated by a local hot-spot in the resonator, i.e the microbridge. The microbridge is assumed to be sufficiently small, so that its temperature $T$ may be considered homogeneous. The temperature of all other parts of the resonator is assumed be equal to that of the coolant $T_{0}$. The heat balance equation for the microbridge reads
\begin{equation}
C\frac{\mathrm{d}T}{\mathrm{d}t}=-H\left(T-T_{0}\right) + 2 \hbar \omega_\mathrm{p} \gamma_{2}\left\vert \hat A \right\vert ^{2}, \label{Eq:dT/dt}
\end{equation}
where $C$ is the thermal heat capacity and $H$ is the heat transfer coefficient.

The coupling mechanism between Eq. (\ref{Eq:dA/dt}) and Eq. (\ref{Eq:dT/dt}) is based on the resonator parameters' dependence on the impedance of the microbridge \cite{Segev_1943}, which is dependent on its phase, i.e., superconductive (SC) vs. normal conductive (NC). We assume the simplest case, where the dependence is a step-function 
at the critical temperature $T=T_c$, namely $\omega_{0},\gamma_{2}$ take values $\omega_{0,\mathrm{s}},\gamma_{2,\mathrm{s}}$ for $T < T_c$ and values $\omega_{0,\mathrm{n}},\gamma_{2,\mathrm{n}}$ for $T > T_C$ \cite{Segev_096206}.

In general, the two coupled Eqs. (\ref{Eq:dA/dt}) and (\ref{Eq:dT/dt}), have two attractor sets \cite{Gurevich_941}: $\lbrace \hat A_{\infty,\mathrm{s}}, T_{\infty,\mathrm{s}} \rbrace$ (the "super attractor") and $\lbrace \hat A_{\infty,\mathrm{n}}, T_{\infty,\mathrm{n}} \rbrace$ (the "normal attractor"). When $T < T_c$ the system evolves toward the super attractor, whereas, when $T > T_c$ it evolves toward the normal attractor. The stability of the super (normal) attractor is dependent on the relation $T_{\infty,\mathrm{s}} < T_c$ ($T_{\infty,\mathrm{n}} > T_c$). Consequently, four stability zones can be identified in the plane of pump power $P_{\mathrm{p}}= \hbar \omega_\mathrm{p} \left\vert a^{\mathrm{in}}\right\vert ^{2}$ vs. pump frequency $\omega_{\mathrm{p}}$ (see Fig. \ref{Fig:StabilityDiagram}).
In the monostable zones, either the SC phase or the NC phase is locally stable; in the bistable zones, both phases are locally stable. In the astable zone, on the other hand, none of the phases are locally stable, and the resonator oscillates between these two phases. As the two phases significantly differ in their reflection coefficients, the oscillations are translated into an amplitude modulation of the reflected pump tone. The borderlines between the four stability zones are given by:\\
\begin{subequations}\label{Eq:Borderlines}
\begin{align}
P_\mathrm{p} &= \frac{H(T_c - T_0)}{4 \gamma_1 \gamma_{2,\mathrm{s}} } \left [ (\omega_\mathrm{p} - \omega_{0,\mathrm{s}})^2 + \gamma_\mathrm{s}^2 \right ], \label{Eq:SuperBorderline}\\
P_\mathrm{p} &= \frac{H(T_c - T_0)}{4 \gamma_1 \gamma_{2,\mathrm{n}} } \left [ (\omega_\mathrm{p} - \omega_{0,\mathrm{n}})^2 + \gamma_\mathrm{n}^2 \right ].
\end{align}
\end{subequations}
Eq. (\ref{Eq:SuperBorderline}) determines the border line between the super conducting mono stable zone and the astable zone which is the modulation threshold. The experiments were done in the vicinity of this borderline (Fig. \ref{Fig:StabilityDiagram}-a).

We now consider an ESR sample under static external magnetic field $\mathbf{B}_z = B_z \hat z$ and MW magnetic field $\mathbf{B}_x = B_x \cos(\omega_\mathrm{p} t) \hat x$, such that $B_x \ll B_z$. Under such conditions the susceptibility of the ESR 
sample is given by \cite{Abragam_NMR,Yalcin_094105}:
\begin{subequations}\label{Eq:ChiPrime}
\begin{align}
\chi^\prime &= -\frac{1}{2}
\frac{(\omega_\mathrm{p} -\omega_\mathrm{l}) t_2^2}
{1 +   (\omega_\mathrm{p} -\omega_\mathrm{l})^2 t_2^2 + \gamma_\mathrm{ESR}^2 B_x^2 t_1 t_2} \omega_\mathrm{l} \chi_0, \\
\chi^{\prime \prime} &= -\frac{1}{2}
\frac{t_2}{1+ (\omega_\mathrm{p} -\omega_\mathrm{l})^2 t_2^2 + \gamma_\mathrm{ESR}^2 B_x^2 t_1 t_2} \omega_\mathrm{l} \chi_0,
\end{align}
\end{subequations}
where $\omega_\mathrm{l} = \gamma_\mathrm{ESR} B_z$ is the Larmor frequency, $t_1$ and $t_2$ are the ESR longitudinal and horizontal relaxation times respectively, and $\chi_0$ is the static susceptibility which is given by $\chi_0 = \mu_0 N \gamma_\mathrm{ESR}^2 \hbar^2 /4k_\mathrm{B} T_\mathrm{ESR}$ for the case of spin 1/2 system with g-factor 2, where 
$T_\mathrm{ESR}$ is the ESR sample temperature.
The interaction with the spins gives rise to changes in the resonator's resonance frequency, $\omega_0$, the dumping coefficient, $\gamma$, and the quality factor, Q, according to the expressions \cite{Poole_ESR}:
\begin{subequations}\label{Eq:DeltaOmegaAndQ}
\begin{align}
\Delta \omega /\omega_\mathrm{0} &\approx \eta \chi ^\prime,\\
\Delta Q /Q_0 \approx -\Delta \gamma &/\gamma \approx - Q_0 \eta \chi ^{\prime \prime},
\end{align}
\end{subequations}
where $\eta$ is the filling factor of the sample \cite{Poole_ESR}. In the linear regime when analytical results are compared with the experimental results in Fig. \ref{Fig:S11} good agreement is found. As can be seen from Eq. (\ref{Eq:SuperBorderline}) the ESR shifts the borderline between the mono-stable and astable zones. Again, when comparing the analytical results with the experimental results in Fig. \ref{Fig:Lockin}, good agreement is 
obtained.

A potential advantage of our methodology can be explained by comparing it to the conventional linear method. In ESR inductive detection with a resonator having a linear response the output signal is relatively small and consequently various readout elements such as cryogenic amplifiers, LIAs, etc. are commonly employed. Reducing the noise of such readout elements to cryogenic temperatures is a major engineering challenge \cite{Rinard_69, Pfenninger_32}. In our scheme, on the other hand, the ESR signal amplification is achieved by the resonator's intrinsic behavior, and consequently no further active amplification is needed. A second potential advantage of our methodology is important for the case of paramagnetic materials with broad resonance lines. This advantage can be understood in relation to LIA measurement and field modulation. Commonly, LIA measurement is applied in order to reduce $1/f$ noise. In the present experiment the LIA detection is carried out with respect to a modulation in the static magnetic field with low frequency. This modulation brings the resonator in/out of the nonlinearity boundary. Such detection scheme is the most common in CW ESR and facilitates its high sensitivity. However, for paramagnetic materials having broad resonance lines reaching 1000 G \cite{Poole_ESR, Misra_EPR} and more, the conventional field modulation at a large amplitude is not possible, as it causes excessive heat and mechanical vibrations. This problem can be solved in our scheme by employing amplitude modulation of the driving MW frequency (instead of field modulation), for which the width of the ESR spectrum does not limit the detection (see Fig. \ref{Fig:StabilityDiagram}).

In summary, we have introduced a new methodology for induction detection of ESR utilizing the non linear response of the MW resonator. We have presented the phenomenon experimentally and theoretically. Our initial
setup is not optimal for ESR measurements of small samples, but in the future, we plan on combining this method with compact MW resonators whose magnetic field is confined to a small volume \cite{Twig_076105}.

We thank Gad Koren and Robert Semerad for advice regarding the fabrication steps. We thank Israel Science Foundation (ISF), German Israel Foundation (GIF), QNEMS STREP, European Research Council (ERC) and RBNI for their financial support.
\bibliography{D:/Dropbox/Technion/Eyal_Bib}

\end{document}